\documentclass[aps,prl,twocolumn,amsmath,amssymb,footinbib,showpacs,longbibliography,superscriptaddress]{revtex4-1}

\usepackage[english]{babel}
\usepackage{latexsym}
\usepackage{graphics}
\usepackage{subfigure}
\usepackage{epsfig}
\usepackage{color}
\usepackage{hyperref}
\usepackage{braket} 
\usepackage[T1]{fontenc}
\usepackage[latin9]{inputenc}
\usepackage{amstext}
\usepackage{amssymb}
\usepackage{graphicx}
\usepackage{esint}
\usepackage{braket}
\usepackage{babel}
\usepackage{amsmath}
\usepackage{siunitx}
\usepackage{ulem}

\newcommand{\Er}{E_{\textrm{r}}}

\newcommand{\kB}{k_{\textrm{\tiny {B}}}}

\newcommand{\rr}{\mathbf{r}}

\newcommand{\asc}{a_{\textrm{\tiny 3D}}}
\newcommand{\aho}{l_\perp}

\newcommand{\gTwoD}{g_{\textrm{\tiny 2D}}}
\newcommand{\gOneD}{g_{\textrm{\tiny 1D}}}
\newcommand{\aOneD}{a_{\textrm{\tiny 1D}}}

\newcommand{\TOneD}{T_{\textrm{\tiny 1D}}}

\newcommand{\Jnature}{Nature (London)}

\newcommand{\Jscience}{Science}

\newcommand{\Jprl}{Phys. Rev. Lett.}

\newcommand{\Jpra}{Phys. Rev. A}

\newcommand{\Jrmp}{Rev. Mod. Phys.}

\begin{document}

\title{Cooling bosons by dimensional reduction}


\author{ Yanliang  Guo }
\thanks{These authors contributed equally to this work.}

\affiliation{Institut f{\"u}r Experimentalphysik und Zentrum f{\"u}r Quantenphysik, Universit{\"a}t Innsbruck, Technikerstra{\ss}e 25, Innsbruck, 6020, Austria}

\author{ Hepeng  Yao }
\thanks{These authors contributed equally to this work.}

\affiliation{DQMP, University of Geneva, 24 Quai Ernest-Ansermet, Geneva, CH-1211 ,  Switzerland}

\author{ Sudipta  Dhar}

\affiliation{Institut f{\"u}r Experimentalphysik und Zentrum f{\"u}r Quantenphysik, Universit{\"a}t Innsbruck, Technikerstra{\ss}e 25, Innsbruck, 6020, Austria}

\author{ Lorenzo  Pizzino}
\affiliation{DQMP, University of Geneva, 24 Quai Ernest-Ansermet, Geneva, CH-1211 ,  Switzerland}

\author{ Milena  Horvath}
\affiliation{Institut f{\"u}r Experimentalphysik und Zentrum f{\"u}r Quantenphysik, Universit{\"a}t Innsbruck, Technikerstra{\ss}e 25, Innsbruck, 6020, Austria}

\author{ Thierry Giamarchi}
\affiliation{DQMP, University of Geneva, 24 Quai Ernest-Ansermet, Geneva, CH-1211 ,  Switzerland}

\author{ Manuele  Landini}
\affiliation{Institut f{\"u}r Experimentalphysik und Zentrum f{\"u}r Quantenphysik, Universit{\"a}t Innsbruck, Technikerstra{\ss}e 25, Innsbruck, 6020, Austria}

\author{ Hanns-Christoph  N{\"a}gerl}\email{christoph.naegerl@uibk.ac.at}
\affiliation{Institut f{\"u}r Experimentalphysik und Zentrum f{\"u}r Quantenphysik, Universit{\"a}t Innsbruck, Technikerstra{\ss}e 25, Innsbruck, 6020, Austria}


\begin{abstract}
Cold atomic gases provide a remarkable testbed to study the physics of interacting many-body quantum systems. They have started to play a major role as quantum simulators, given the high degree of control that is possible. A crucial element is given by the necessarily non-zero temperature. However cooling to the required ultralow temperatures or even simply measuring the temperature directly on the system can prove to be very challenging tasks. Here, we implement thermometry on strongly interacting two- and one-dimensional Bose gases with high sensitivity in the nano-Kelvin temperature range. Our method is aided by the fact that the decay of the first-order correlation function is very sensitive to the temperature when interactions are strong. We find that there may be a significant temperature variation when the three-dimensional quantum gas is cut into two-dimensional slices or into one-dimensional tubes. Strikingly, the temperature for the one-dimensional case can be much lower than the initial temperature. Our findings show that this decrease results from the interplay of dimensional reduction and strong interactions.
\end{abstract}

\maketitle
\def\thefootnote{*}\footnotetext{These authors contributed equally to this work}\def\thefootnote{\arabic{footnote}}

Cold atomic gases allow a remarkable degree of control over crucial parameters such as the interaction strength and the confining potentials, making them ideal systems for studying the properties of strongly correlated quantum matter~\cite{bloch-review-2008} . Their dimensionality can be set freely, via, e.g., optical lattice potentials, and with this they have enabled the study of a host of properties of interacting one-dimensional (1D) and two-dimensional (2D) quantum systems. Highlights in 1D include the observation of bosonic fermionization into the Tonks-Girardeau (TG) state~\cite{paredes2004,kinoshita2004,haller2009,meinert-bloch-2017}, the driving of quench dynamics~\cite{Hofferberth2007,meinert-quench-2014,Weiss-fermion2020,Weiss-prethermal-2023}, the investigation into localization effects driven by longitudinal lattices ~\cite{stoferle2004,clement-1d-2009,haller2010,fabbri-1d-2012,boeris-1dshallow-lattice-2016} and disorder~\cite{billy2008,derrico2014,gori2016,Schneider-2DBG-2020}, and the recent observation of spin-charge separation~\cite{vijayan-spin-charge-2020,Hulet-spin-2022}. Similarly, 2D systems based on cold atoms have allowed the study of the Berezinskii-Kosterlitz-Thouless (BKT) transition~\cite{hadzibabic2006,ha-strong2Dboson-2013}, the investigation of topological properties~\cite{goldman-topology-2013,tarnowski-topology-2019}, and the probing of frustrated phases~\cite{struck2011}. In all these works, it has been important to assure that the temperature is low enough compared to the point of quantum degeneracy as non-zero temperatures play a crucial role for the physical phenomena observed.

The presumably lowest temperatures for 2D and 1D systems have so far been achieved by slicing low-entropy 3D samples such as essentially pure atomic Bose-Einstein condensates (BEC) into layers or tubes by means of lattice potentials. While the temperatures of the 3D sources can be determined to high accuracy, with values in the low-nK range \cite{bloch-review-2008}, estimates of the temperatures of the low-D systems have always been rather vague. In fact, it is widely believed that creating low-D gases within anisotropic potentials usually leads to heating~\cite{fabbri-1D-T-2011,hadzibabic2006,gori2016,li-1D-T-2020}. Attempts to obtain a temperature value from e.g. Bragg-spectroscopy data on 1D Luttinger liquids together with exact Bethe-ansatz modeling \cite{meinert-1Dexcitation-2015} were hampered by rather large systematic uncertainties. Here, we implement precise thermometry at the 1-nK level for strongly-interacting 2D and 1D Bose gases. We utilize the fact that the first-order correlation function $g^{(1)}$ sensitively depends on temperature when interactions are strong. In the experiment, it is determined from a careful measurement of the momentum distribution, and the results of ab-initio state-of-the-art quantum Monte Carlo calculations for various values of the temperature are used as a thermometer scale. We use the thermometer to determine temperatures in 1D that are significantly lower than the starting temperatures.  We are able to interpret this anomalous phenomenon by invoking an entropy argument and find that our data fits well with the theoretical prediction. We attribute our findings to the interplay of tight confinement and strong interactions for bosons that are subject to fermionization.

\begin{figure*}[t!] 
\centering
\includegraphics[width=1.5\columnwidth]{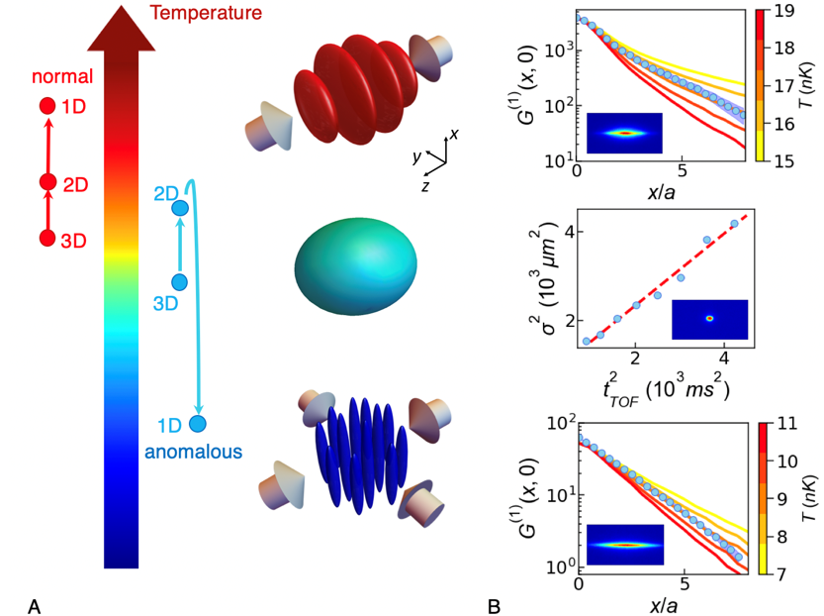}
\caption{\label{fig:sketch-thermometer} 
\textbf{Sketch of the experimental setup and typical thermometry data for the various dimensions.}
(A) The initial nearly spherical 3D BEC (center) is cut either into an ensemble of 2D layers (top) or into an array of 1D tubes (bottom) via the optical force of one or two pairs of counter-propagating and interfering laser beams (arrows). The temperature scale illustrates the normal and anomalous temperature change when the dimensionality is switched for two slightly different initial conditions. (B) Example data for the temperature measurements in 2D (top), 3D (middle), and 1D (bottom). For 2D and 1D, the calculated one-body correlation function $G^{(1)}(x,0)$ is plotted as a function of distance $x/a$ for various temperatures (solid lines, with the temperature indicated by the color coding) and compared to the measured data (blue circles). The experimental statistical error from 20 repetitions is smaller than the size of the symbols. For 2D case, the system has a weighted atom number $\overline{N}_{\textrm{\tiny 2D}}\sim3950$ with radial trapping frequency $\omega^{\textrm{\tiny 2D}}_x/2\pi=10.1$\SI{}{\hertz}. For 1D case, the system has a weighted atom number $\overline{N}_{\textrm{\tiny 1D}}\sim63$ with longitudinal trapping frequency $\omega^{\textrm{\tiny 1D}}_x/2\pi=14.3$\SI{}{\hertz}. The scattering length is set to $a_{\textrm{\tiny 3D}}=620a_0$ for both 2D and 1D. The QMC calculations are under the experimental conditions and its error bars are less than $1\%$. For the 3D case, a typical TOF dataset with the squared Gaussian waist $\sigma^2$ obtained from a bimodal fit as a function of the squared TOF duration $t_\text{TOF}$ is presented. The linear fit (dashed line) directly gives the 3D temperature. The three insets are example TOF absorption images for the respective dimensionality~\cite{supplemental}.
}
\end{figure*}

%

The experiment starts with an interaction-tunable 3D Bose-Einstein condensate (BEC) of $1.5\!\times\!10^5$ Cs atoms~\cite{Kraemer2004} prepared in the lowest magnetic hyperfine state $\vert F,m_F \rangle=\vert 3,3 \rangle$, held in a crossed-beam dipole trap with trap frequencies $\omega_{x,y,z}\!=\!2\pi\times(18.6(2), 19.3(3), 26.8(3))\SI{}{\hertz}$ along the three main axes $x, y$, and $z$ of the setup and levitated along the vertical $x$-direction against gravity by a magnetic field gradient. The BEC is in the Thomas-Fermi regime with the 3D $s$-wave scattering length $a_{\textrm{\tiny 3D}}$ tuned to $a_{\textrm{\tiny 3D}}\!\approx\!190 a_0$. One (or two) counterpropagating optical lattice beams along the $z$- (and $y$-) direction are gradually ramped up in \SI{500}{\milli\second} to a potential depth of $V_z=30E_r$ ($V_y\!=\!30\Er$), with $\Er\!=\! \pi^2\hbar^2/(2ma^2)$ the recoil energy, cutting the 3D system into an ensemble of 2D layers that lie in the $x$-$y$-plane (or an array of 1D tubes along the $x$-direction), as sketched in Fig.~\ref{fig:sketch-thermometer}A. Here, $a\!=\!\lambda/2$ is the lattice spacing with $\lambda=1064.5$ nm the wavelength of the lattice light. After loading the atoms into the lattice, the initial crossed-beam dipole trap is ramped down in \SI{100}{\milli\second}. For the layers, the trap frequencies are $\omega_{x,y,z}\!=\!2\pi\times(10.1(2), 10.1(2), 11{\rm k})\SI{}{\hertz}$, and these change to $\omega_{x,y,z}\!=\!2\pi\times(14.3(2), 11{\rm k}, 11{\rm k})\SI{}{\hertz}$ for the set of 1D tubes.  The offset magnetic field is then ramped up adiabatically to set $a_{\textrm{\tiny 3D}}\!\approx\!620 a_0$. This takes the 3D BEC into the strictly-2D regime with 2D interaction parameter $\gamma_{\textrm{\tiny 2D}}\!\!=\!\!1.5$ or into the strictly-1D regime with Lieb-Linger parameter $\gamma_{\textrm{\tiny 1D}}\!=\!20$~\cite{supplemental}. For this value, the 1D system is deep in the fermionized TG regime \cite{kinoshita2004,haller2009}.

\begin{figure*}[t!] 
\centering
\includegraphics[width=1.6\columnwidth]{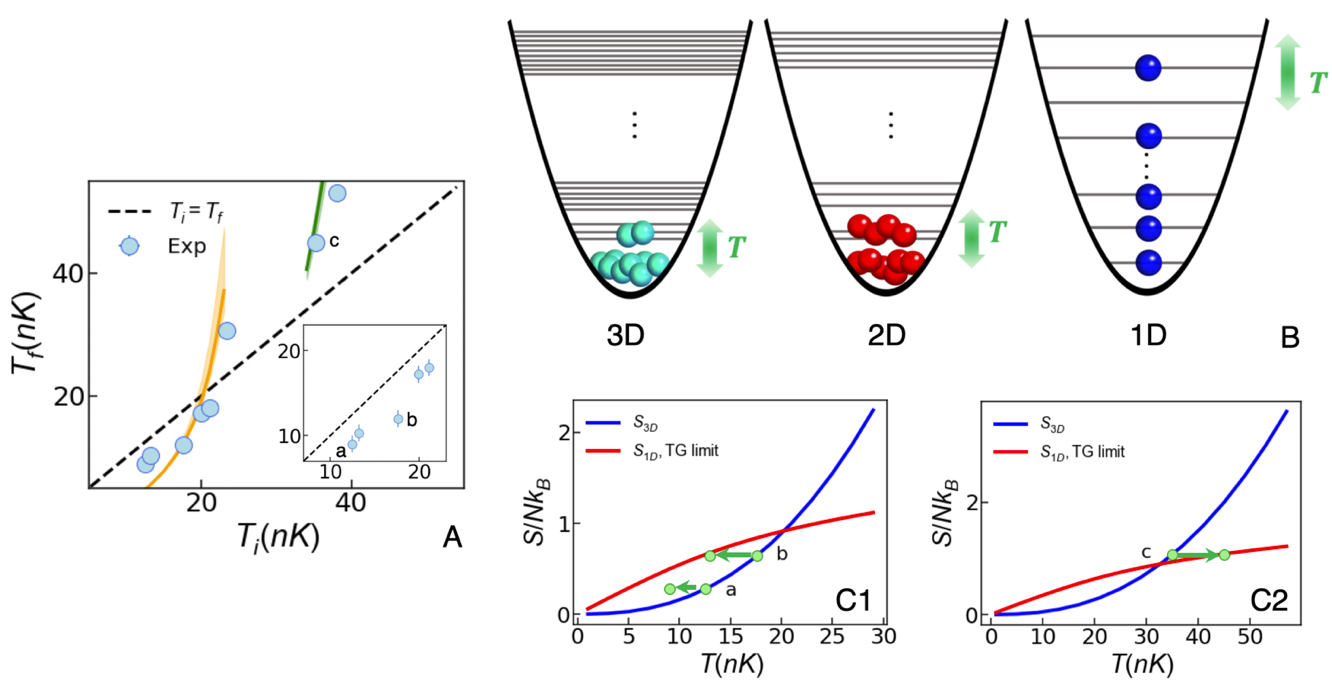}
\caption{\label{fig:entropy} 
\textbf{Cooling vs. heating and the physical picture behind anomalous cooling.} (A) The final temperature of the 1D system $T_f$ (blue circles) as a function of the initial temperature of the 3D system $T_i$. The error bars are smaller than the size of the symbols. The analytical predictions are shown as orange and green solid curves. For this data, the 1D systems are always deeply in the TG regime. The inset displays the low-temperature data for which cooling is observed. The letters a, b, and c mark data points that are referenced in C1 and C2. (B) Illustration of the configuration picture in 3D, 2D, and 1D on the basis of the quantum harmonic oscillator. The green arrows indicate the effect of the non-zero temperatures. (C1-C2) The entropy per particle $S/N\kB$ for the 3D trapped system (blue line) and the 1D tubes deeply in the TG regime (red line) as a function of the temperature $T$. The experimental parameters are $\omega_{x,y,z}=2\pi\!\times\!(18.6, 19.3, 26.8)\SI{}{\hertz}$ and $\overline{N}_{\textrm{\tiny 1D}}=72$ for the data in C1 and $\omega_{x,y,z}=2\pi\!\times\!(29.4, 27.1, 39.9)\SI{}{\hertz}$ and $\overline{N}_{\textrm{\tiny 1D}}=120$ for the data in C2. The green dots reflect the three cases shown in (A), among which a and b show the cooling effect, while c shows heating. 
}
\end{figure*}

We now detail the thermometer operation in the various dimensionalities. The method for 3D is standard and has been used widely in the past. In short, the temperature is determined from the expansion rate of the non-condensed fraction of the quantum gas in time-of-flight (TOF) after switching off all trapping fields. Crucially, the inter-particle interaction is zeroed at the start of TOF by means of a Feshbach resonance's zero crossing \cite{Kraemer2004} to avoid any residual interaction effects. Bimodal fits on the density profiles for varying TOF-times give the 3D temperature~\cite{supplemental}. Typically, $8$ different TOF-times are chosen, and each time the experiment is repeated 2 times. Fig.~\ref{fig:sketch-thermometer}B provides the results of a typical measurement, for which we find $T_{\textrm{\tiny 3D}}=12.5(4) \SI{}{\nano\kelvin}$. In 2D and 1D, the interacting gases do not show a bimodal distribution and a Boltzmann fit cannot be done. However, the one-body correlation function $g^{(1)}(x,x',y,y')=\langle \hat{\Psi}^\dagger(x', y')\hat{\Psi}(x, y)\rangle$ shows a decay that has a strong temperature dependence. In the experiment, we determine it by a measurement of the momentum distribution $n(k)$ via the TOF technique to obtain the integrated correlation function $G^{(1)}(x,y)=\iint dx^{\prime}dy^{\prime} g^{(1)}(x^{\prime}+x,y^{\prime}+y,x^{\prime},y^{\prime})$ via Fourier transform. We then compare it to the results of an ab-initio quantum Monte Carlo (QMC) approach to simulate the system \cite{yao-crossoverD-2022}. Its many-body Hamiltonian is given by  
\begin{equation}\label{eq:Hamiltonian}
\hat{H} = \sum_j \left[ -\frac{\hbar^2}{2m} \nabla^2_j + V(\hat{\rr}_j) \right] + \sum_{j<k} U(\hat{\rr}_j - \hat{\rr}_k),
\end{equation}
with $U(\hat{\rr})$ the short-range repulsive two-body interaction and $V(\rr) $ the external harmonic potential. The function $G^{(1)}(x,y)$ is then computed for various temperatures using the worm algorithm. Note that simulating one weighted tube (layer) gives us the same result for $G^{(1)}(x,y)$ as taking into account the whole atom distribution in the array of tubes (layers) \cite{supplemental}. Fig.~\ref{fig:sketch-thermometer} B presents typical experimental data for $G^{(1)}(x,0)$ in 1D and 2D and compares the data to the results of the QMC simulations for various temperatures. Clearly, the QMC data serves as a very sensitive ruler for the temperature. For 2D, we obtain $T_{\textrm{\tiny 2D}}=17(1) \ \SI{}{\nano\kelvin} $, and the temperature in 1D is $T_{\textrm{\tiny 1D}}=9(1)\ \SI{}{\nano\kelvin}$. Evidently, the system is hotter in 2D, and then colder in 1D. The 1D data can be cross-checked using the analytical form of the correlation function. For a trapped 1D TG gas it reads $G^{(1)}(x)\sim e^{-\eta x/a}$ with $\eta=\kB T a/ (2\hbar^2 n_0)$~\cite{minguzzi-boson-exact-2022}. With $n_0\!=\!0.9/a$ and $\eta_\text{exp}=0.48$, we find $T_{\textrm{\tiny 1D}}^\text{analytical}\!=\!9.1\ \SI{}{\nano\kelvin}$. This agrees well with the QMC prediction.

The data above was taken for a specific set of parameters. We now perform cross-checks by varying the initial 3D temperature, the trapping frequency, and the interaction strength to elucidate the mechanism behind the anomalous cooling. For example, by changing the efficiency of the evaporative cooling process in the initial 3D dipole trap, we prepare 3D quantum-degenerate samples at various initial temperatures $T_i$ with varying condensate fractions~\cite{supplemental}. These samples are then transferred into 1D tubes and we measure the final temperature $T_f$ as before. Such temperature data is shown in Fig.~\ref{fig:entropy}A. Clearly, anomalous cooling occurs when the initial temperatures are sufficiently low, i.e., 20 nK and below. Typcially, we see a decrease of $20\%$ to $40\%$ from the initial 3D temperature, with a temperature difference that is far more than 1-nK thermometer resolution. However, above $T_i \approx 20$ nK the dimensional change leads to heating.

\begin{figure*}[t!] 
\centering
\includegraphics[width=1.6\columnwidth]{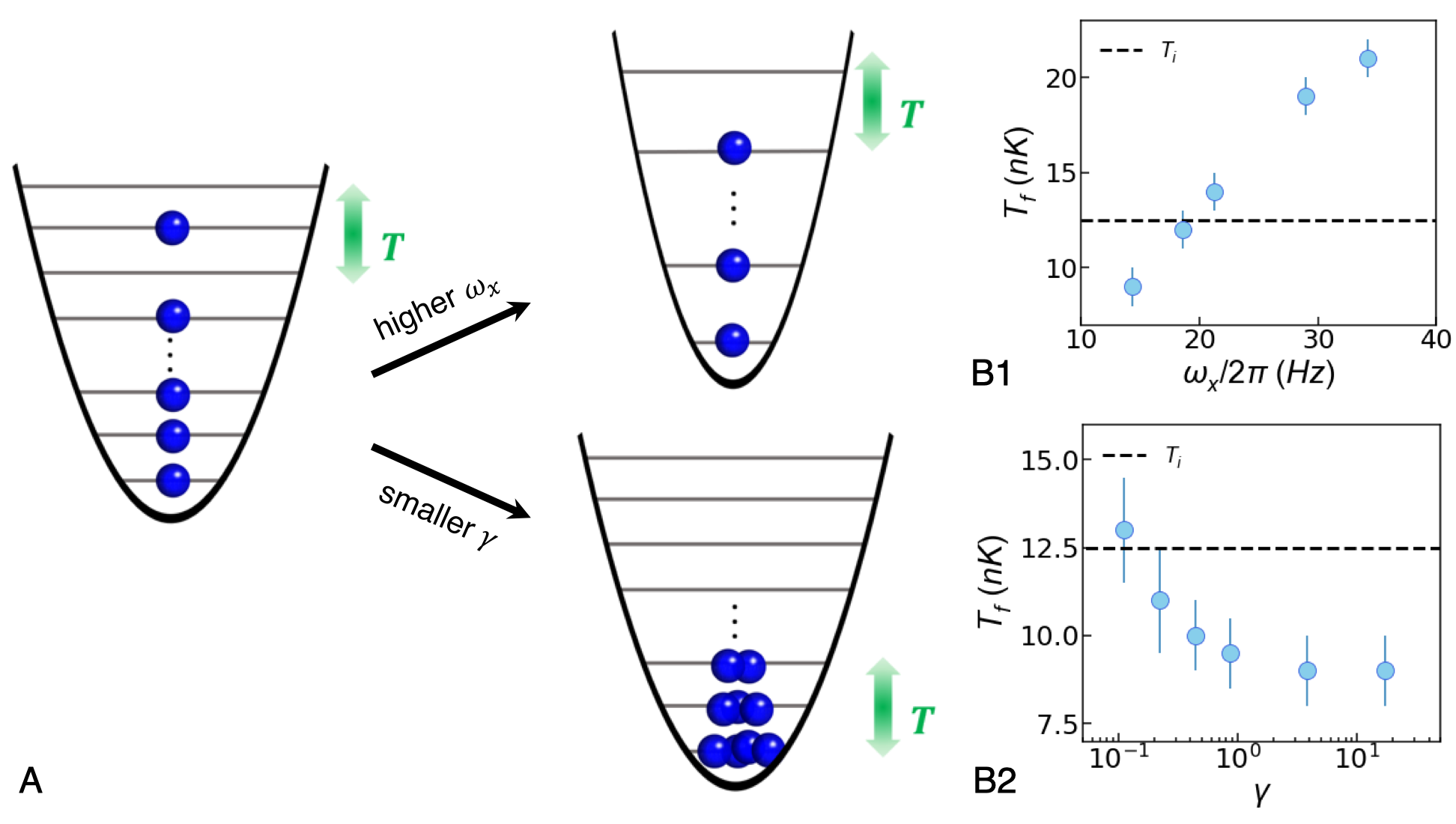}
\caption{\label{fig:condition} 
\textbf{Conditions for anomalous cooling.} (A) Illustration of the configuration picture for the fermionized TG gas, comparing to the one with a higher trapping frequency $\omega_x$ or a weaker interaction strength. (B1-B2) The measured 1D temperature $T_{f}$ as a function of the longitudinal trapping frequency $\omega_x$ (B1) and the interaction strength $\gamma$ (B2). The black dashed line indicates the initial 3D temperature $T_i$. }
\end{figure*}

Invoking the entropy picture sheds light onto the anomalous cooling phenomenon and demonstrates the important role played by the dimensionality of the quantum many-body system. Fig.~\ref{fig:entropy}B  illustrates the population of the energy levels of the quantum harmonic oscillator in the different dimensionalities. When the dimension of the system is reduced from 3D, two processes happen. On one hand, the condensate nature of the initial system is undermined. In 3D, the system is a nearly pure BEC with a small non-condensed fraction at low temperatures. Most of the atoms populate the ground state. In 2D, the nature of condensate is weakened and the system exists only as a quasicondensate with a decay of the first-order correlations. This suggests an increase of the number of possible configurations $C$ in energy space for a given non-zero temperature $T$. In the extreme case of a strongly-interacting gas in 1D, the system has fermionized. The particles are filled into the energy levels as ideal fermions and the excitations happen around the Fermi surface. On the other hand, the degeneracy of the energy levels becomes less as the dimensionality is reduced. This leads to the decrease of the number of possible configurations $C$. Thus, as a result of the competition of these two processes, one can reach a situation $C_{\textrm{\tiny 1D}}>C_{\textrm{\tiny 3D}}>C_{\textrm{\tiny 2D}}$. For constant entropy, as a result of careful adiabatic loading the lattice, one may thus obtain $T_{\textrm{\tiny 1D}}<T_{\textrm{\tiny 3D}}<T_{\textrm{\tiny 2D}}$.

This physical picture is confirmed by calculations of the entropy. For 3D trapped bosons~\cite{zhou-cooling-lattice-2007}, the entropy is $S_{\textrm{\tiny 3D}}=(7A \zeta(3)/5\sqrt{2}) (15 \asc N / \sigma )^{1/5}(T/\hbar \bar{\omega})^{5/2}$ with $A=10.6$, $\sigma=\sqrt{\hbar/m\bar{\omega}}$
the oscillator length and $\bar{\omega}=(\omega_x \omega_y \omega_z)^{1/3}$. In the 1D case, the entropy $S_{\textrm{\tiny 1D}}=-\partial\ \Omega_{\mathrm{TG}}/\partial\ T$ of a TG gas can be computed from the grand potential  $\Omega_{\mathrm{TG}}$~\cite{xu2015}, with the trap treated under the local density approximation ~\cite{supplemental}. These entropy curves are shown in Fig.~\ref{fig:entropy}C for our set of parameters. Below a certain temperature $T^c_i$, the entropy for 1D is higher than in 3D. When keeping the entropy constant, the system's temperature has to drop when the dimensionality is reduced from 3D to 1D, see e.g. a and b in Fig.~\ref{fig:entropy}A. Our data reflects this. Above $T^c_i$, a temperature increase is expected. This is also captured by our data. In Fig.~\ref{fig:entropy}A, we add the prediction for the temperature decrease resp. increase during dimensional reduction. Our data fits the predictions reasonably well. Only for the lowest temperatures we find a decrease that is not as pronounced as predicted, most likely due to some small non-adiabaticity during the lattice-loading process. We note that the cooling mechanism observed here is reminiscent of the adiabatic demagnetization cooling technique~\cite{cooling1933,pfau2006}. However, in our case there is no discrete spin degree of freedom into which entropy can be pumped.

We next turn to the influence of a change of the longitudinal trapping frequency $\omega_{x}$ for the 1D systems. By means of the crossed-beam trap, we can tune $\omega_{x}/2\pi$ from $14.3(2)$ Hz to $34.2(3)$ Hz. Our data, shown in Fig.~\ref{fig:condition}B1, shows that the 1D temperature then varies from $9$ nK to $21$ nK. This confirms our entropy and configuration picture: stiffening of the confinement reduces the number of accessible configurations, as shown schematically in Fig.~\ref{fig:condition}A, and hence leads to an increase of the temperature. As is well know, adiabatic compression of a Boltzmann gas leads to heating, and decompression leads to cooling. Indeed, our data shows that also a TG gas has the same behavior. We note that the lower limit $14.3$ Hz for $\omega_{x}$ is set by the residual transversal trapping force of the $y$- and $z$-lattice beams. Reducing this value would require some anti-trapping, which could be done by means of an additional blue-detuned laser beam, and with this even lower 1D temperatures should be possible.

We finally address the role of strong interactions. In the experiment, after preparing the 1D tubes, we ramp $a_{\textrm{\tiny 3D}}$ to a value between 7$a_0$ and 620$a_0$, varying $\gamma$ between $0.1$ and $20$ by more than 2 orders of magnitude given our typical atom number $\bar{N}$ \cite{supplemental}. We find a clear temperature dependence on $\gamma$ as seen in Fig.~\ref{fig:condition}B2~\cite{supplemental}. As $\gamma$ is increased, the 1D temperature drops continuously. Above $\gamma\!\approx\!1$ the temperature settles to a constant value. Evidently, more configurations become accessible as the system starts to fermionize, as sketched in Fig.~\ref{fig:condition}A, leading to a reduction of the temperature, and beyond $\gamma_{\textrm{\tiny 1D}}\!\approx\!1$ the system's fermionization is complete for a system in equilibration.

In conclusion, we have realized a thermometer for strongly-interacting 1D and 2D quantum gases. We are capable of measuring temperatures for such strongly-correlated systems in the low-nK range with 1-nK precision. With this thermometer, we have found that cooling may occur as the dimensionality is reduced from 3D to 1D. Given such a thermometer, one can now optimize the formation process of the low-D quantum gases, in particular for the case of non-harmonic box-like trapping conditions~\cite{Navon2021}. For example, a detailed probing of the 2D-to-1D crossover regime is possible \cite{guo-crossoverD-2023}. The principle demonstrated here could be used to generate even lower temperatures in a Carnot-cycle process. Next, a variety of phenomena in low-D becomes accessible for which the temperature plays an important role, such as Anderson localization~\cite{billy2008}, the pinning~\cite{haller2010} and Bose glass transitions~\cite{derrico2014,Schneider-2DBG-2020}, and out-of-equilibrium dynamics with e.g. pre-thermalization~\cite{Weiss-prethermal-2023}, dynamical fermionization~\cite{Weiss-fermion2020}, correlated transport~\cite{flutter2012}, and the implementation of quantum-field machines~\cite{jorg-prx}.

\acknowledgments
The Innsbruck team acknowledges funding by a Wittgenstein prize grant under project number Z336-N36 and by the European Research Council (ERC) under project number 789017. This research was funded in part by the Austrian Science Fund (FWF) W1259-N27 and MH thanks the doctoral school ALM for hospitality. This work is also supported by the Swiss National Science Foundation under grant number 200020-188687. The numerical calculations make use of the ALPS scheduler library and statistical analysis tools~\cite{troyer1998,ALPS2007,ALPS2011}.

The data that support the findings of this study are made publicly available from Zenodo by the authors at \cite{Zenodo}.

\bigskip

\section*{Supplemental material}

\subsection*{The atom number distribution}
\label{app:weighted-N}

\begin{figure}[t!] 
\centering
\includegraphics[width=1.0\columnwidth]{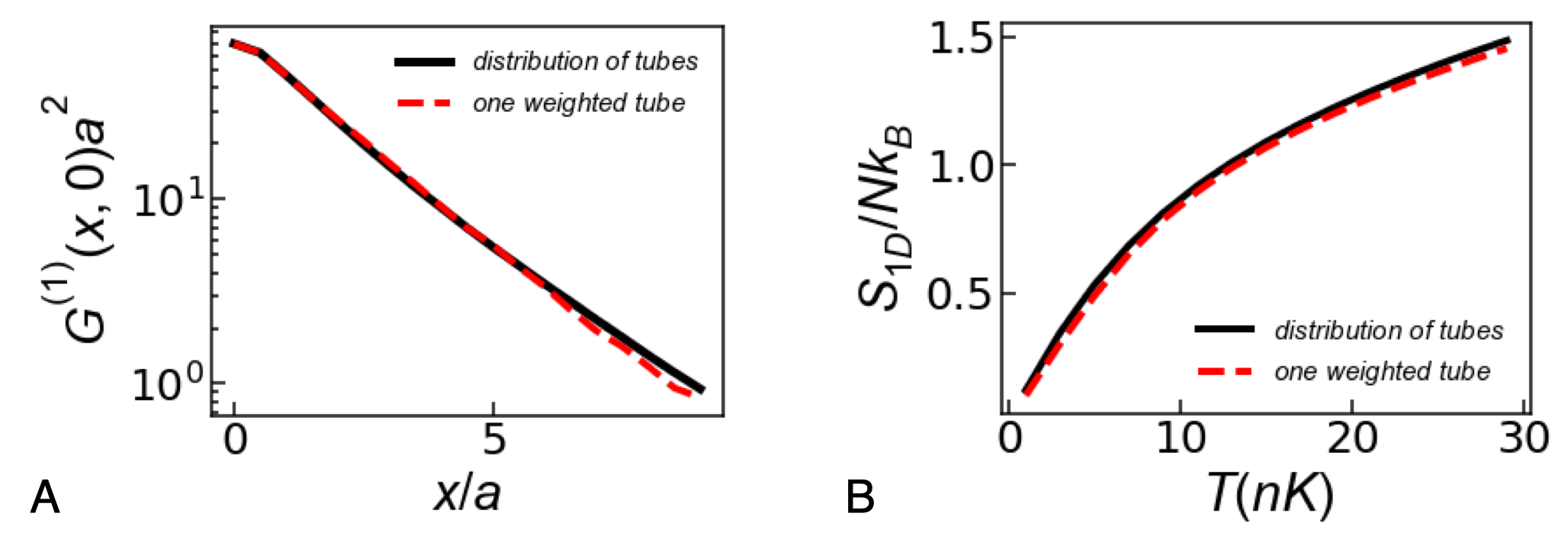}
\caption{\label{fig:supp-distribution} 
\textbf{Check that $\overline{N}$ is a proper input for the numerical calculations.}  (A) The correlation function $G^{(1)}(x,0)$ as a function of the position $x/a$ for the same parameters as for the 1D experimental data shown in Fig.~\ref{fig:sketch-thermometer}B of the main text and for $T=9$ nK. (B) The entropy per particle $S/N \kB$  as a function of temperature $T$ for a TG gas with $\bar{N}=40$ and $\omega_{x}=2\pi\times 14.3$ Hz. The red dashed line indicates the calculation for the case of one weighted tube, while the black solid line assumes the distribution of tubes. 
}
\end{figure}

For the calculations of this paper, we use one single layer or one single tube with a weighted average number $\bar{N}$ given by
\begin{equation}\label{eq:weighted-N}
\bar{N}= \frac{\sum_{i} N_{i}^2}{\sum_{i} N_{i}}
\end{equation}
with $N_i$ the atom number of the $i$-th tube (or layer). This method uses the atom number as a weight of itself and it takes into account that  the tube with more atoms will contribute more significantly to the physical properties of the systems than tubes with less atoms. Such a standard method has been used in previous works ~\cite{Haller-2011,meinert-1Dexcitation-2015,li-1D-T-2020}. During the lattice-loading process almost all layers (2D) or tubes (1D) are in the Thomas-Fermi regime for weak repulsive interactions, so we can calculate the initial occupation number in each layer in 2D or each tube in 1D case through the global chemical potential and the total atom number. In the following, we check that this method is proper for the physical quantities studied here. First, we compute the correlation function $G^{(1)}(x,0)$ for the 1D systems as done for the QMC calculations shown in Fig.~\ref{fig:sketch-thermometer}B of the main text, see Fig.~\ref{fig:supp-distribution} A. The results for one single weighted tube agree with the calculations for the full tube distribution. The difference is less than $5\%$, showing that the use of the weighted average atom number is good enough for estimating the temperature in low dimensions. Second, we check that also the result for the entropy using one weighted tube agrees with the calculation using the distribution of tubes. The results are shown Fig.~\ref{fig:supp-distribution} B for a TG gas under the condition $\bar{N}=40$ and $\omega_x\!=\!2\pi\times 14.3$ Hz. Also here we find good agreement.

The interaction strengths in 1D and 2D are then determined by the using the weighted atom number. In 1D, the coupling constant is \cite{bloch-review-2008,cazalilla2011}
\begin{equation}\label{eq:coupling-g1d}
\gOneD= \frac{2 \hbar^2 \asc}{m\aho^2} \bigg(1-\frac{1.036\asc}{\aho}\bigg)^{-1}
\end{equation}
with $\aho= \sqrt{\hbar/m\omega_\perp}$ the characteristic transverse length, and the 1D Lieb-Liniger parameter is hence given by $\gamma_{\textrm{\tiny 1D}}=m \gOneD /\hbar^2\bar{n}a$ with the density $\bar{n}$ of the weighted average tube. In 2D, the coupling constant writes \cite{bloch-review-2008,hadzibabic-2Dgas-2011}
\begin{equation}\label{eq:coupling-g2d}
\gTwoD \simeq \frac{2\hbar^2 \sqrt{2\pi}}{m \aho/\asc+{1/\sqrt{2\pi} \ln (1/\pi q^2 \aho^2)}}
\end{equation}
with $q=\sqrt{2m\vert\mu\vert/\hbar^2}$ the quasi-momentum, and $\mu$ the chemical potential corresponding to the weighted average layer. The 2D interaction parameter thus is $\gamma_{\textrm{\tiny 2D}}=m \gTwoD /\hbar^2$.

\subsection*{Thermometer in the weakly-interacting regime}
\label{app:weak-thermometer}

\begin{figure}[t!] 
\centering
\includegraphics[width=0.85\columnwidth]{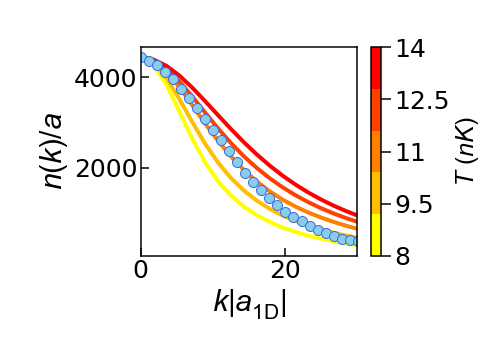}
\caption{\label{fig:supp-nk-weak} 
\textbf{Thermometer for 1D weakly-interacting systems based on the momentum distribution.} The blue balls are experimental data with weighted atom number $\overline{N}_{\textrm{\tiny 1D}}\sim70$, longitudinal trapping frequency $\omega_x/2\pi=14.3$ \SI{}{\hertz} and 1D coupling constant $m\gOneD a/\hbar^2=0.11$. The momentum axis is rescaled by $a_{\textrm{\tiny 1D}}\!\!=\!\!-2\hbar^2/m\gOneD$. The solid lines are the results of QMC simulations for the same conditions as the experiment for various values of the temperature as indicated.
}
\end{figure}

For most of the cases in the main text, we estimate the temperature of the system by using the one-body correlation function $G^{(1)}(x,0)$. In the strongly interacting regime, this quantity follows a purely exponential decay and its decay exponent has a strong and clear dependence on temperature. However, in the weakly interacting regime, the correlation decay becomes slower and it is not a purely exponential decay, which makes the analysis more complicated. In this case, we extract the temperature of the system directly from the momentum distribution $n(k_x)$. This method has been used to estimate the temperature of 1D weakly-interacting gases in previous experimental studies~\cite{Gerbier-1D-T-2003,fabbri-1D-T-2011,yangbing2017}.
In Fig.~\ref{fig:supp-nk-weak}, we present one example for $\overline{N}_{\textrm{\tiny 1D}}\sim70$, longitudinal trapping frequency $\omega_x/2\pi=14.3$ \SI{}{\hertz} and 1D coupling constant $m\gOneD a/\hbar^2=0.11$, resulting in $\gamma\sim0.2$. Our QMC simulations provide a thermometer scale with a temperature sensitivity of $\delta T\approx\SI{1.5}{nK},$ see the colored solid curves. With the data from the range $k\aOneD<20$, the measurement suggests 1D temperature $\TOneD=\SI{11}{nK}\pm\SI{1.5}{nK}$. Notably, such an estimate for weakly-interacting gases according to the small-$k$ part of the momentum distribution is similar to the one mentioned in Refs.~\cite{Gerbier-1D-T-2003,fabbri-1D-T-2011}. 
Instead of using the analytical formula for the width of the distribution, our calculation directly simulates the continuous trapped system and avoids the local density approximation.We notice that the resolution of the thermometer in the regime of weak interactions is not quite as good as the resolution the regime of strong interactions. This is due to the fact that for weak interaction the system is less sensitive to the temperature due to the more pronounced quantum coherence.

\subsection*{The preparation of different initial 3D temperatures}
\label{app:experiment}

\begin{figure}[t!] 
\centering
\includegraphics[width=0.65\columnwidth]{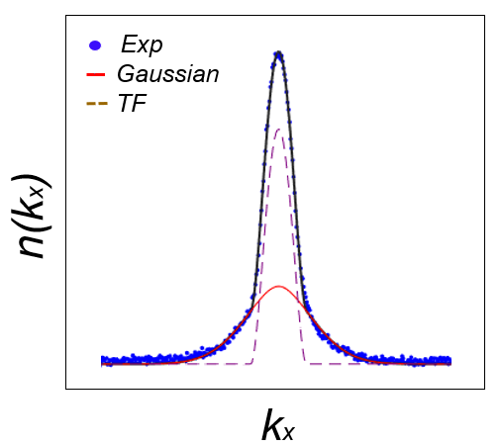}
\caption{\label{fig:supp-nk} 
\textbf{ Bimodal fit for the momentum distribution $n(k_x)$ in 3D.} Example of the $n(k_x)$ for the 3D gas under a $50$-ms TOF in Fig.~\ref{fig:sketch-thermometer} (B) in the main text with a bimodal fit.}

\end{figure}

In Fig.~\ref{fig:entropy} A of the main text we show how the final temperature of the 1D system varies with the initial 3D temperature. In the experiment, we are able to prepare a 3D gas at different temperatures by changing the efficiency of the evaporative cooling process, i.e., varying the trap depth through altering the power of the crossed dipole beam. 
However, once the trap depth is modified, the 3D trapping frequency and the weighted atom number in 1D are changed consequently, resulting in different entropy for both cases in 3D and 1D, as shown in Fig~\ref{fig:entropy} C. To systematically compare the measured 1D temperature to the theoretical predicted value that drawn as the orange and the green solid lines in Fig~\ref{fig:entropy}A , we have to keep the 3D trapping frequency constant. So after evaporative cooling, we first ramp up the power of the crossed dipole beam to a fixed value, which provides a constant 3D trapping frequency to $\omega_{\textrm{\tiny 3D}}=2\pi\times(18.6, 19.3, 26.8)\SI{}{\hertz}$ and then load atoms into the lattice.
However, such a trap depth limits the reachable highest temperature for a 3D gas. For preparing an even hotter 3D cloud, we need to increase the power of the crossed-dipole beams setting the final trapping frequency $\omega_{\textrm{\tiny 3D}}=2\pi\times(29.4, 27.1, 39.9)\SI{}{\hertz}$. This allows us to measure the 1D temperatures to the predicted value given by the entropy curves in  Fig~\ref{fig:entropy} C1 and C2.

Moreover, we argue that the process of measuring the 3D temperature is accurate.
The 3D temperatures in the main text are all measured after ramping up the crossed dipole beam for the desired $\omega_{\textrm{\tiny 3D}}$. 
In addition, the inter-particle interaction is switched off during TOF expansion. The resulting true momentum distribution $n(k_x)$ shows an obverse thermal part that can be perfectly fitted by a bimodal (Gaussian plus Thomas-Fermi), see Fig~\ref{fig:supp-nk}. 
It help us to extract the 3D temperature accurately as shown in Fig.~\ref{fig:sketch-thermometer}B.

\subsection*{The imaging setup}

\begin{figure}[h] 
\centering
\includegraphics[width=0.95\columnwidth]{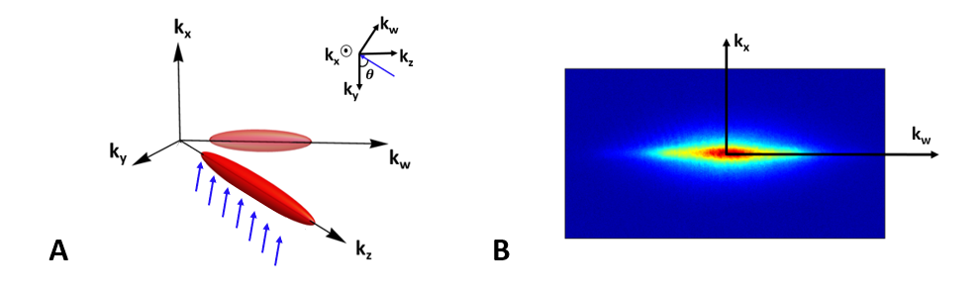}
\caption{\label{fig:sketch111}
\textbf{A schematic of our imaging setup and an example of the image.} (A) The vectors $k_y$, $k_z$, $k_w$ and the imaging direction (blue arrows) all lie in one plane, with $k_x$ perpendicular to this plane, see also inset. The red 3D ellipsoid along the $k_z$ direction indicates the atomic cloud after TOF starting from an ensemble of 2D layers for $V_{y} = 0 E_{r}$. The light red 2D ellipsoid along the $k_w$ direction is the shadow in our absorption image. (B) The projected image after TOF for the 2D case. }

\end{figure}

Our setup consists of two lattice beams perpendicular to each other along the $y$ and $z$ axes. There is an angle of $\theta \sim 57^\circ$  between the propagation axis of the imaging beam and the $y$-direction, see Fig. \ref{fig:sketch111}. However, $x$ axis is always perpendicular to the imaging beam. We note that, to determine the temperature in any dimensions, we capture our momentum distributions $n(k_x)$ or the correlation functions $G^{(1)}(x,0)$ along the longitudinal direction $x$ in the main text, which is not affected by this angle.

\subsection*{The entropy of the TG gas}
\label{app:entropy-1d}

The entropy of the TG gas in an harmonic trap can be computed based on the equations in Ref.~\cite{xu2015}. The grand potential density $\Omega$ of a homogeneous TG system can be expressed to the lowest order as
\begin{equation}\label{eq:grand-tonks}
\Omega (\mu, T)= - \frac{\sqrt{m} (\kB T)^{3/2}}{\hbar \sqrt{2\pi}} f_{3/2} (\mu/\kB T)
\end{equation}
with $\mu$ the chemical potential and $f_{\nu} (x)$ the completed Fermi-Dirac integral at index $j$.  
Correspondingly, the entropy density $s$ of this point can be obtained by 
\begin{equation}\label{eq:entropy-1D}
s(\mu, T)=-\frac{\partial \Omega(\mu,T)}{\partial T} \bigg\rvert_{\mu}.
\end{equation}
Then, the presence of the harmonic trap can be taken care of by using the local density approximation (LDA) by taking the equation of state for the low temperature TG gas, namely
\begin{equation}\label{eq:eos-tonks}
n=\sqrt{\frac{2m\mu}{\pi^2 \hbar^2}}.
\end{equation}
This means that the entropy of the trapped system can be expressed as 
\begin{equation}\label{eq:grand-tonks-2}
S_\text{trap}(N,T)=\int_{-R}^{+R}\ dx\ s(\mu(x),T)
\end{equation}
under the constrain of the total number of particles $\int_{-R}^{+R}\ dx\ n(\mu-1/2 m\omega^2x^2)=N$.


\begin{thebibliography}{10}
\bibitem{Zenodo}
{Data set is available from {Z}enodo at doi: 10.5281/zenodo.8212006}.
\end{thebibliography}
\end{document}